\documentclass[twocolumn,floats,floatfix,superscriptaddress,aps,pra]{revtex4}
\usepackage{amsfonts}
\usepackage{amssymb}
\usepackage{amsmath}
\usepackage{calc}
\usepackage{graphicx}
\usepackage{bm}
\usepackage{makecell}
\usepackage{booktabs}

\def\be{ \begin{equation} }
\def\ee{ \end{equation} }
\def\bea{ \begin{eqnarray} }
\def\eea{ \end{eqnarray} }

\def\bse{ \begin{subequations} }
\def\ese{ \end{subequations} }
\def\ba{ \begin{array} }
\def\ea{ \end{array} }

\def\i{\,\text{i}}
\def\e{\,\text{e}}
\def\i{i}
\def\e{e}

\def\fromto{\leftrightarrow}

\def\U{\mathbf{U}}
\def\F{\mathbf{F}}

\def\H{\mathbf{H}}

\def\i{{\rm{i}}}

\def\be{ \begin{equation} }
\def\ee{ \end{equation} }
\def\bea{ \begin{eqnarray} }
\def\eea{ \end{eqnarray} }
\def\bse{ \begin{subequations} }
\def\ese{ \end{subequations} }

\def\X{\mathcal{X}}

\def\U{\mathbf{U}}
\def\B{\mathcal{B}}
\def\BN{\mathcal{B}_N}
\def\C{\mathcal{C}}

\def\S{\mathcal{S}}

\def\F{\mathcal{F}}
\def\H{\mathbf{H}}

\def\i{i}
\def\e{e}

\def\fromto{\leftrightarrow}

\def\subsec#1{\textbf{#1}}

\def\ket#1{\vert #1 \rangle}
\def\bra#1{\langle #1 \vert}
\def\Qa{Q^{0}}
\def\Qb{Q^{1}}
\def\Q{\mathbf{Q}}
\def\Ra{R^{0}}
\def\Rb{R^{1}}
\def\S{\mathbf{S}}
\def\R{\mathbf{R}}

\def\b{\xi}

\begin{document}

\author{Boyan T. Torosov}
\affiliation{Institute of Solid State Physics, Bulgarian Academy of Sciences, 72 Tsarigradsko chauss\'{e}e, 1784 Sofia, Bulgaria}
\author{Nikolay V. Vitanov}
\affiliation{Department of Physics, St Kliment Ohridski University of Sofia, 5 James Bourchier blvd, 1164 Sofia, Bulgaria}

\title{High-fidelity composite quantum gates for Raman qubits}

\date{\today}

\begin{abstract}
We present a general systematic approach to design robust and high-fidelity quantum logic gates with Raman qubits using the technique of composite pulses.
We use two mathematical tools --- the Morris-Shore and Majorana decompositions --- to reduce the three-state Raman system to an equivalent two-state system.
They allow us to exploit the numerous composite pulses designed for two-state systems by extending them to Raman qubits.
We construct the NOT, Hadamard, and rotation gates by means of the Morris-Shore transformation with the same uniform approach: sequences of pulses with the same phases for each gate but different ratios of Raman couplings.
The phase gate is constructed by using the Majorana decomposition.
All composite Raman gates feature very high fidelity, beyond the quantum computation benchmark values, and significant robustness to experimental errors.
All composite phases and pulse areas are given by analytical formulas, which makes the method scalable to any desired accuracy and robustness to errors.
\end{abstract}

\maketitle

%%%%%%%%%%%%%%%%%%%%%%%%%%%%%%%%%%%%%%%%%%%%%%%%%%%%%%%%%%%%%%%%%%%%%%%%%%%%%%%%%%%%%%%%%%%%%%%%%%%%%%%%%%%%%%%%%%%%%%%%%%%%%%%%%%%%%%%%%
%%%%%%%%%%%%%%%%%%%%%%%%%%%%%%%%%%%%%%%%%%%%%%%%%%%%%%%%%%%%%%%%%%%%%%%%%%%%%%%%%%%%%%%%%%%%%%%%%%%%%%%%%%%%%%%%%%%%%%%%%%%%%%%%%%%%%%%%%
%%%%%%%%%%%%%%%%%%%%%%%%%%%%%%%%%%%%%%%%%%%%%%%%%%%%%%%%%%%%%%%%%%%%%%%%%%%%%%%%%%%%%%%%%%%%%%%%%%%%%%%%%%%%%%%%%%%%%%%%%%%%%%%%%%%%%%%%%
%\sec{Introduction\label{Sec:intro}}
%
Composite pulses (CPs) --- sequences of pulses with well defined relative phases --- have enjoyed tremendous success as a basic control tool of simple quantum systems over the last 40 years.
Developed in nuclear magnetic resonance (NMR) \cite{NMR} they have spread due to their unique features to many other fields, including trapped ions \cite{Gulde2003,Schmidt-Kaler2003,Haffner2008,Timoney2008,Monz2009,Shappert2013,Mount2015,Vitanov2015,Randall2018,Sriarunothai2019}, neutral atoms  \cite{Rakreungdet2009}, doped solids \cite{Schraft2013,Genov2017,Bruns2018,Genov2020}, optical atomic clocks \cite{Zanon-Willette2018}, cold-atoms interferometry \cite{Butts2013,Dunning2014,Berg2015}, optically dense atomic ensembles \cite{Demeter2016}, quantum dots \cite{Wang2012,Kestner2013,Wang2014,Zhang2017,Hickman2013,Eng2015}, NV centers in diamond \cite{Rong2015}, magnetometry \cite{Aiello2013},
optomechanics \cite{Ventura2019}, etc.
Recently, many new types of CPs have been developed in order to boost the fidelity of some well known quantum control techniques, e.g. rapid adiabatic passage \cite{Torosov2011PRL,Schraft2013}, stimulated Raman adiabatic passage \cite{Torosov2013,Bruns2018}, Ramsey interferometry \cite{Vitanov2015,Zanon-Willette2018}, and dynamical decoupling \cite{Genov2017}.

Composite pulse sequences feature a unique combination of ultrahigh fidelity similar to resonant excitation and robustness to experimental errors similar to adiabatic techniques.
Moreover, CPs offer a great flexibility unseen in other control technique: they can produce broadband (BB), narrowband (NB), passband (PB), and virtually any desired excitation profile.
These features render CPs ideal for applications in quantum computation and quantum technologies in general \cite{QIP}.

Quantum technologies use qubits which are implemented either as a directly or indirectly coupled two-state quantum system.
For example, in trapped ions, the electronic states of the ions are used as qubits of two types: optical and radio-frequency (rf) qubits.
Optical qubits consist of an electronic ground state and a metastable state, with lifetimes of the order of seconds, while the rf qubits are usually encoded in the hyperfine levels of the electronic ground states of the ion, with lifetimes of thousands of years.
Either of these come with their advantages and disadvantages.
It has been shown that by using dressing fields in the rf-qubits configuration, one can suppress decoherence, caused by magnetic-field fluctuations, by as many as three order of magnitude \cite{DressedQubit, DressedQubit2}.
Rf qubits can be manipulated directly \cite{Wunderlich}, by Raman transitions \cite{Monroe}, or by combinations of these \cite{DressedQubit, DressedQubit2}.
For a directly coupled qubit, just two states suffice and CPs are implemented directly to them.
For indirectly coupled qubit states, via an ancillary middle state, CPs are scarce, if any, because their construction requires to control more complicated multistate quantum dynamics.

In this Letter, stimulated by the advances described above, we develop a general systematic approach for creating robust and high-fidelity quantum gates in Raman-type qubits.
Our method is based on the use of composite pulse sequences, adapted for a three-state system.
While the vast amount of literature on composite pulses has been focused on two-state systems, studies of CPs in higher-dimensional systems also exist \cite{MultiLevelCPs, MultiLevelCPs1, MultiLevelCPs2, Randall2018, MultiLevelCPs4, GenovMultistate,Ivanov2015-2,Ivanov2015-34,Jones2013}.
Our method uses two powerful mathematical techniques: the Morris-Shore transformation \cite{MS} and the Majorana decomposition \cite{Majorana}, which map the three-state Raman system onto equivalent two-state systems.
Below we briefly introduce these two techniques and then design the single-qubit gates used in quantum computing.

%%%%%%%%%%%%%%%%%%%%%%%%%%%%%%%%%%%%%%%%%%%%%%%%%%%%%%%%%%%%%%%%%%%%%%%%%%%%%%%%%%%
%%%%%%%%%%%%%%%%%%%%%%%%%%%%%%%%%%%%%%%%%%%%%%%%%%%%%%%%%%%%%%%%%%%%%%%%%%%%%%%%%%%
%%%%%%%%%%%%%%%%%%%%%%%%%%%%%%%%%%%%%%%%%%%%%%%%%%%%%%%%%%%%%%%%%%%%%%%%%%%%%%%%%%%

\begin{figure}[tb]
	\includegraphics[width=8cm]{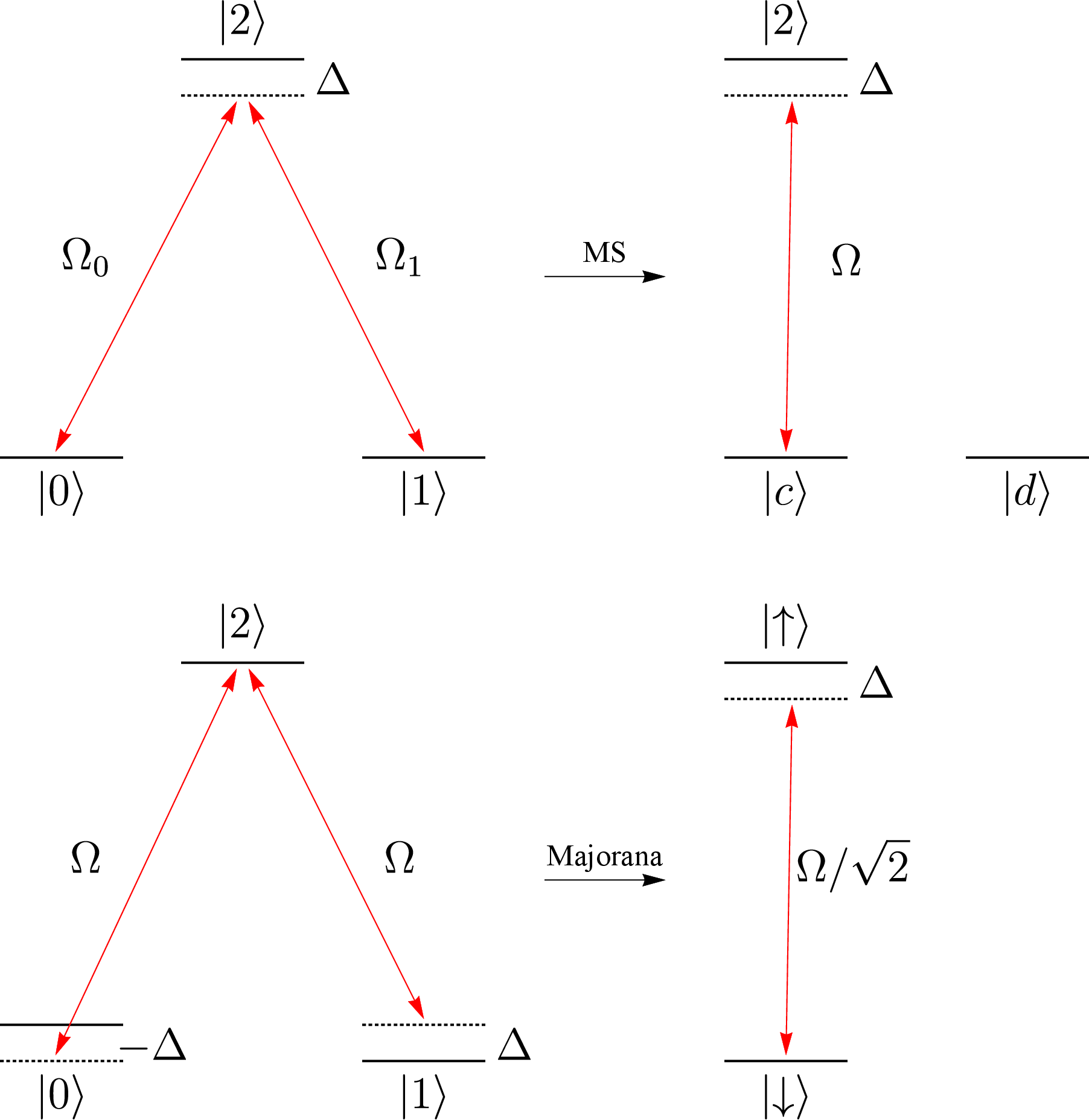}
	\caption{
		(Left) Lambda system, representing a Raman qubit, consisting of states $\ket{0}$ and $\ket{1}$ coupled to an excited state $\ket{2}$. 
(Right) After the Morris-Shore (top) or Majorana (bottom)  transformations, the Raman system reduces to an effective two-state problem.
	}
	\label{fig:scheme}
\end{figure}

We assume that the Raman qubit consists of two ground states, $\ket{0}$ and $\ket{1}$, coupled to an excited state $\ket{2}$, as illustrated in Fig.~\ref{fig:scheme} (left).
Both the Morris-Shore transformation \cite{MS}, and the Majorana decomposition \cite{Majorana} allow one to reduce the three-state Raman system to a two-state problem, Fig.~\ref{fig:scheme} (right).
We note that when the one-photon detuning in Fig.~\ref{fig:scheme} (top left) is large, we can adiabatically eliminate state $\ket{2}$ and obtain an effective two-state system $\{ \ket{0}, \ket{1} \}$.
However, the large detuning reduces the effective coupling $\ket{0} \fromto \ket{1}$ and increases the gate time.
The Morris-Shore and Majorana decomposition work for any detuning, and even on resonance, when the dynamics, and the gates, are fastest.

%=================================================================
%=================================================================

\subsec{Morris-Shore (MS) transformation.}
It requires the Hamiltonian of the $\Lambda$ system to have the form
$\H(t)=\frac12 \hbar [ \Omega_0(t) e^{i\phi_0} \ket{0} \bra{2} + \Omega_1(t) e^{i\phi_1} \ket{1} \bra{2} + \text{h.c.} ] + \hbar\Delta \ket{2} \bra{2} $,
where $\Delta$ is the single-photon frequency detuning and $\Omega_{0,1}(t)$ are the Rabi frequencies of the two transitions,
 which must share the same time dependence, $\Omega_{0}=\b_0 f(t)$ and $\Omega_{1}=\b_1 f(t)$.
%For our particular objective here, based on numerical evidence, 
We assume that the two phases are equal, $\phi_0=\phi_1 \equiv \phi$.
We introduce the root-mean-square (RMS) Rabi frequency $\Omega(t)=\sqrt{\Omega_{0}(t)^2 + \Omega_{1}(t)^2} = \b f(t)$, where $\b=\sqrt{\b_0^2+\b_1^2}$.
Without loss of generality, we assume that $\int f(t)dt=\pi$.
The two pulse areas are $A_k = \b_k \pi$ ($k=0,1$), and the RMS pulse area is $A=\b\pi$.
The MS transformation decomposes the three-state system into a decoupled (dark) state $\ket{d} = (\b_1 \ket{0} - \b_0 \ket{1}) / \b $ and a two-state system, consisting of a state $\ket{c} = (\b_0 \ket{0} + \b_1 \ket{1}) / \b $ coupled to the excited state $\ket{2}$, as shown in Fig.~\ref{fig:scheme} (top right) \cite{MS, GenovMultistate}.
The Hamiltonian in the MS basis is $\H_{\text{MS}}(t) = \frac12{\hbar} [ \Omega(t) e^{i \phi} \ket{c} \bra{2} + \text{h.c.} ] + \hbar\Delta \ket{2} \bra{2} $.
The corresponding MS propagator in the basis $\{ \ket{d}, \ket{c}, \ket{2} \}$ in the most general form reads
\be\label{prop-ms}
\U_{\text{MS}}=\e^{-\i\delta}\left[ \begin{array}{ccc}
	\e^{\i\delta} & 0 & 0 \\
	0 & a & b e^{i\phi}  \\
	0 & -b^{\ast} e^{-i\phi} & a^{\ast}
\end{array}\right],
\ee
where $a$ and $b$ are complex Cayley-Klein parameters and $\delta=\int\Delta dt/2$ is a phase factor coming from the representation of the Hamiltonian.
For resonant excitation, we have $a = \cos(A/2)$ and $b = -\i\sin(A/2)$.
The propagator in the original basis reads
\be\label{prop-original}
\U(\phi) = \frac{\e^{-\i\delta}}{\b^2}\left[ \begin{array}{ccc}
	a\b_0^2+\e^{\i\delta}\b_1^2 & \b_0\b_1 (a-\e^{\i\delta}) & \b_0 \b b e^{i \phi} \\
	\b_0\b_1 (a-\e^{\i\delta}) & \e^{\i\delta}\b_0^2+a\b_1^2 & \b_1 \b b e^{i \phi} \\
	-\b_0 \b b^{\ast} e^{-i \phi} & -\b_1\b b^{\ast} e^{-i \phi} & \b^2 a^{\ast}
\end{array}\right].
\ee
If we apply a sequence of such pairs of pulses, each with some relative phase $\phi_k$, we can use the phases as free parameters to construct composite Raman pulses.

%If we impose a common phase shift to both Rabi frequencies, $\Omega_{0,1}\to \Omega_{0,1}\e^{\i\phi}$, this phase shift is mapped on the RMS Rabi frequency, $\Omega \to \Omega\e^{\i\phi}$.
%This property is used to construct composite Raman pulses.

\subsec{Majorana decomposition.}
The Majorana decomposition reduces a multistate system with the SU(2) symmetry to a two-state problem.
Explicitly, it maps the three-state Hamiltonian
$\H(t) = \frac12 \hbar [ \Omega(t)\e^{\i\phi} \ket{0} \bra{2} + \Omega(t)\e^{-\i\phi} \ket{1} \bra{2} + \text{h.c.} ] + \hbar\Delta ( \ket{1} \bra{1} - \ket{0} \bra{0} ) $ onto the two-state Hamiltonian
$\H_M(t) = \frac{1}{2\sqrt{2}} \hbar [ \Omega(t)\e^{\i\phi} \ket{\!\downarrow} \bra{\uparrow\!} + \text{h.c.} ] + \frac12\hbar \Delta ( \ket{\!\uparrow} \bra{\uparrow\!} - \ket{\!\downarrow} \bra{\downarrow\!})$ \cite{Majorana, GenovMultistate, Randall2018}.
If the two-state propagator is parameterized as
\be\label{prop-2s}
\U_M = \left[ \begin{array}{cc}
	a & b e^{i \phi} \\
	-b^{\ast} e^{-i \phi} & a^{\ast}
\end{array}\right],
\ee
then the three-state propagator is
\be\label{prop-3s}
\U(\phi) = \left[ \begin{array}{ccc}
	a^2 & b^2 e^{2i \phi} & \sqrt{2} a b e^{i \phi} \\
	b^{\ast 2} e^{-2i \phi} & a^{\ast 2} & -\sqrt{2} a^{\ast} b^{\ast} e^{-i \phi} \\
	-\sqrt{2} a b^{\ast} e^{-i \phi} & \sqrt{2} a^{\ast} b e^{i \phi} & |a|^2 - |b|^2
\end{array}\right].
\ee
%\textbf{NV: fix notation for propagators!\\}
We shall use this mapping to design high-fidelity composite Raman gates.

%%%%%%%%%%%%%%%%%%%%%%%%%%%%%%%%%%%%%%%%%%%%%%%%%%%%%%%%%%%%%%%%
%%%%%%%%%%%%%%%%%%%%%%%%%%%%%%%%%%%%%%%%%%%%%%%%%%%%%%%%%%%%%%%%
%%%%%%%%%%%%%%%%%%%%%%%%%%%%%%%%%%%%%%%%%%%%%%%%%%%%%%%%%%%%%%%%
%\sec{Composite Raman gates.} %\label{Sec-QIP}
%
%In this section we show how one can produce some of the most basic operations in quantum computing, using the Raman-qubit framework, described in the previous section. Namely, we consider

We are now ready to construct composite Raman implementations of the basic single-qubit quantum gates:
 the $X$, Hadamard, rotation, and phase-shift gates.

\subsec{$X$ gate.} %\label{sec-piPulse}
%
%We start with the simplest operation, used in QIP: an $\X$ gate.
The $X$ gate is defined as the Pauli's matrix $\hat{X}=\hat{\sigma}_x = \ket{1}\bra{0} + \ket{0}\bra{1}$, and it is the quantum equivalent of the classical NOT gate.
%This operation can be achieved by a resonant $\pi$-pulse. %which inverts the populations of the qubit states.
%, using a resonant pulse with an area, equal to $\pi$.
%
One way to produce the Raman $X$ gate, as seen from Eq.~\eqref{prop-original}, is to choose the Rabi frequency amplitudes as $\b_0=-\b_1=\sqrt{2}$.
Then $a=-1$ and $b=0$, and the propagator %in the original basis
 reads
\be\label{pi-pulse-prop}
\U=\left[ \begin{array}{ccc}
	0 & 1 & 0 \\
   1 &  0 & 0 \\
	0 &  0 & -1
\end{array}\right],
\ee
which is the $X$ gate for the qubit $\{\ket{0},\ket{1}\}$.
This operation, however, suffers from the drawbacks of resonant excitation: errors in the experimental parameters (Rabi frequencies, pulse durations,  detuning) reduce the fidelity.
% For instance, instead of the desired Rabi frequencies $\Omega_{0,1}$, we may have $\Omega_{0,1}(1+\epsilon)$, where $\epsilon$ is some (hopefully small) deviation, leading to errors.

The composite pulses overcome these drawbacks.
%Instead of a single pulse, one has a sequence of pulses with well-defined relative phases, which are used as control parameters to shape the excitation profile in any desired way.
For Raman transitions, instead of a single pair of pulses, we use a sequence of $N$ pulses pairs with well-defined relative phases.
%, which are used as control parameters to shape the excitation profile in any desired way.
%
The overall propagator reads
\be\label{CP}
\U^{(N)} =\U(\phi_N)\U(\phi_{N-1})\cdots \U(\phi_2)\U(\phi_1),
\ee
where $\U(\phi)$ is the propagator %of the $\Lambda$ system
 for a single pulse pair, Eq.~\eqref{prop-original} or Eq.~\eqref{prop-3s}.
% after a common phase shift $\phi$ has been applied to both Rabi frequencies, $\Omega_k\to \Omega_k\e^{\i\phi}$ $(k=0,1)$.
We measure the performance of the $X$ gate in the figures below by the infidelity $D =\sqrt{ \sum_{jk} \left|X_{jk} - U^{(N)}_{jk}\right|^2 }$, defined as the distance between the target gate $X$ and the actual propagator $\U^{(N)}$.

\emph{\textbf{Morris-Shore.}}
As it is evident from Eqs.~\eqref{prop-ms} and \eqref{prop-original}, a composite sequence in the original basis transforms into a composite sequence in the MS basis.
This feature allows us to use the vast library of composite pulses in two-state systems to design Raman composite pulses.
We note that the $\pi$ pulse \eqref{pi-pulse-prop} in the original basis corresponds to a $2\pi$ pulse in the MS basis.
Hence, our goal is to obtain a robust $2\pi$ pulse in the MS basis, which will map onto a robust $\pi$ pulse in the original basis.

%=================================================================
\begin{figure}[tb]
	\includegraphics[width=7.cm]{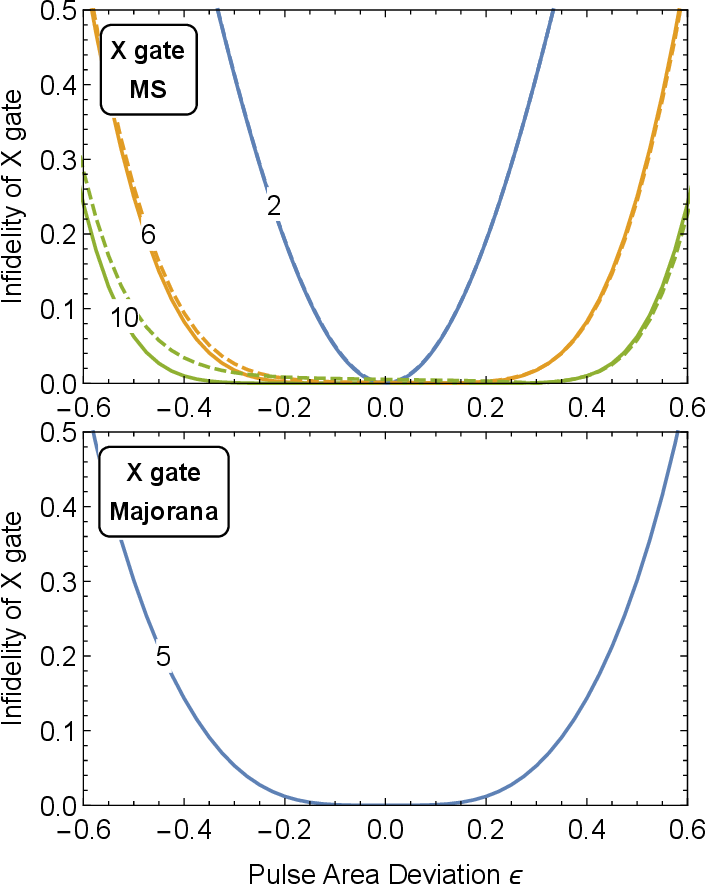}
	\caption{
		(Top) Infidelity of the $X$ gate as a function of the pulse area error for a composite sequence of two, six, and ten pairs of pulses ($N=1,3,5$) by using the MS transformation. The explicit sequences for $N=3$ and $N=5$ are given by Eqs.~\eqref{Raman-BB3BB3}. The dashed curves show the corresponding infidelities in the presence of a small detuning, $\Delta T = 0.1$  [See Eq.~\eqref{Raman-BB3BB3-detuning}]. (Bottom) Same as top, but using the Majorana decomposition and the pulse sequence \eqref{Majorana-X}.
	}
	\label{fig:PiPulseVsArea}
\end{figure}
%=================================================================

$2\pi$ CPs are not as ubiquitous in the literature as $\pi$ CPs.
We propose here to create a $2\pi$ CP by merging two broadband (BB) $\pi$ CPs $\BN$, each consisting of $N$ pulses,
\be\label{twopi}
\C_{2N} = \BN \BN ,
\ee
where $\BN$ are the BB composite sequences \cite{Torosov2011}
\bse\label{BBCP}
\begin{align}
\BN &= B_{\phi_1} B_{\phi_2}\ldots B_{\phi_N} , \label{BBCP} \\
\phi_k &= \frac{k(k-1)}{N}\pi,\quad (k=1,\ldots,N). \label{phases}
\end{align}
\ese
Here $B = \pi (1+\epsilon)$ is a nominal $\pi$ pulse (i.e., for error $\epsilon=0$).
For $N=3$, we have the famous sequence $\B_3 = B_0 B_{2\pi/3} B_0$ and hence $\C_6 = \B_3 \B_3 = B_0 B_{2\pi/3} B_0 B_0 B_{2\pi/3} B_0 $.
The CP \eqref{twopi} features error compensation in both the populations \emph{and} the phases of the propagator, which makes it suitable for gates.
This is because this type of broadband $2\pi$-pulses is in fact a special case of the phase-gate composite pulses, derived earlier \cite{PhaseGate}.
This sequence applies to the MS basis.
To obtain each of the nominal $\pi$ pulses, we choose $\b_{0}=-\b_{1}=\frac{1}{\sqrt{2}}$.
Therefore, each $\pi_{\phi_k}$ pulse in the MS basis corresponds to the pulse pair $(\Qa_{\phi_k}, -\Qb_{\phi_k})\equiv \Q_{\phi_k}$ in the original basis, where $Q^{0,1} = \frac{\pi}{\sqrt{2}} (1+\epsilon)$ are nominal $\pi/\sqrt{2}$ pulses, and $\phi_k$ corresponds to the same phase in the two fields.
Therefore, the first two BB composite Raman sequences read%
\bse\label{Raman-BB3BB3}
\begin{align}
\X_{6} &= \Q_0 \Q_{\frac23\pi} \Q_0 \Q_0 \Q_{\frac23\pi} \Q_0, \label{Raman-BB3BB3a}\\ \label{Raman-BB3BB3b}
\X_{10} &= \Q_0 \Q_{\frac25\pi} \Q_{\frac65\pi} \Q_{\frac25\pi} \Q_0 \Q_0 \Q_{\frac25\pi} \Q_{\frac65\pi} \Q_{\frac25\pi} \Q_0.
\end{align}
\ese
The infidelity of the resulting $X$ gate for such sequences is shown in Fig.~\ref{fig:PiPulseVsArea} (top, solid lines).

\textbf{\emph{Nonzero detuning.}}
When one-photon detuning is present, as illustrated in Fig.~\ref{fig:scheme} (top), we can proceed in the following way.
Instead of $a=-1$, as in the resonant case, now we need to obtain $a=-\e^{\i\delta}$, as seen from the propagator \eqref{prop-original}.
This can be done by producing a phase gate $ \hat{F} = \exp[i \eta \sigma_z] $,  with a phase $\eta = \pi+\delta $.
A robust composite phase gate can be produced by a sequence of two BB $\pi$ CPs, the first one with a zero phase and the second with a phase $\eta$ \cite{PhaseGate}.

If the detuning is \emph{small}, $|\Delta| \ll \Omega,1/T$, where $T$ is the pulse width, we can replace the sequence of two broadband CPs \eqref{twopi} with phased CPs,
 \be\label{phasedtwopi}
 \C_{2N} = \BN(0)\BN(\delta),
 \ee
and in such way obtain an approximation to the phase gate $\hat{F}$.
%In this way, it is straightforward to show by expanding the Cayley-Klein parameters in the Rabi model, that we obtain approximately the desired result \eqref{Xgate}.
Explicitly, the total sequence for $N=3$, analogous to the sequence \eqref{Raman-BB3BB3a}, is
%\be
%(\pi)_0(\pi)_{2\pi/3}(\pi)_0(\pi)_{\delta/2}(\pi)_{\delta/2+2\pi/3}(\pi)_{\delta/2} .
%\ee
\be\label{Raman-BB3BB3-detuning}
%(\Qa_0 \Qb_0) (\Qa_{\frac23\pi} \Qb_{\frac23\pi}) (\Qa_0 \Qb_0) (\Qa_\xi \Qb_\xi) (\Qa_{\frac23\pi+\xi} \Qb_{\frac23\pi+\xi}) (\Qa_\xi \Qb_\xi) .
\X_{6} = \Q_0 \Q_{\frac23\pi} \Q_0 \Q_{\delta} \Q_{\frac23\pi+\delta} \Q_{\delta}.
\ee
%Therefore, the same composite sequences can be used in the presence of small detuning as in the resonant case, only with minor modification.
The infidelity of the $X$ gate for such sequences is shown in Fig.~\ref{fig:PiPulseVsArea} (top, dashed lines).

%=================================================================
\begin{figure}[tb]
	\includegraphics[width=7.cm]{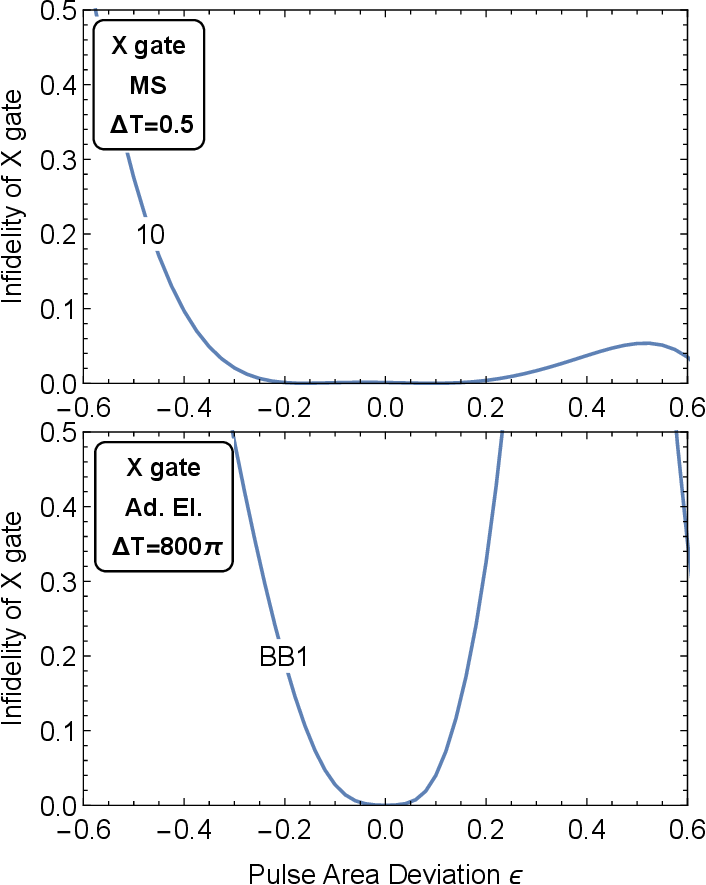}
	\caption{
		Infidelity of the $X$ gate as a function of the pulse area error for: (top) composite sequence of ten pairs of pulses ($N=5$), obtained by the sequence \eqref{Raman-universal} of two universal CPs, in the presence of moderate detuning and (bottom) BB1 composite sequence \eqref{BB1} in the presence of large detuning. The adiabatic elimination in the bottom frame uses $\Omega_{0}=\Omega_{1}=40\pi/T$.
	}
	\label{fig:PiPulseVsAreaDetuning}
\end{figure}
%=================================================================

If the detuning is \emph{moderate}, $|\Delta| \sim \Omega,1/T$, one can use CPs with double compensation in the pulse area and the detuning, and produce the $X$ gate as in Eq.~\eqref{phasedtwopi}.
For instance, the five-pulse universal CP \cite{GenovUniversal}
%\be\label{U5}
$\mathcal{U}_5 = B_0 B_{\frac56\pi} B_{\frac13\pi} B_{\frac56\pi} B_0$
%\ee
produces a composite $X$ gate in the presence of moderate detuning by applying the sequence
\begin{align}\label{Raman-universal}
\X_{10} = \Q_0 \Q_{\frac56\pi} \Q_{\frac13\pi} \Q_{\frac56\pi} \Q_0
\Q_{\delta} \Q_{\frac56\pi+\delta} \Q_{\frac13\pi+\delta} \Q_{\frac56\pi+\delta} \Q_{\delta},
\end{align}
The performance is illustrated in Fig.~\ref{fig:PiPulseVsAreaDetuning} (top frame).

Finally, if the detuning is \emph{large}, $|\Delta| \gg \Omega,1/T$, one can adiabatically eliminate the excited state $\ket{2}$ and obtain an effective two-state system with the effective two-photon coupling $\Omega_{\text{eff}}=-\Omega_{0}\Omega_{1}^\ast/(2\Delta)$
\footnote{A more accurate approach for elimination of state $\ket{2}$, which is applicable not only for large but even for moderate and small detuning, uses the effect of adiabatic population return \cite{Torosov2012}.}.
We can directly apply CPs in this system.
Therefore, by applying a composite $\pi$-pulse with phase stabilisation, we achieve an $X$ gate, up to a global phase, due to the present Stark shift after the adiabatic elimination.
One prominent example of such pulse sequence is the BB1 composite pulse \cite{Wimperis1994},
\be\label{BB1}
\text{BB}1=B_{\zeta} B_{3\zeta} B_{3\zeta} B_{\zeta} B_{0},
\ee
where $\zeta=\arccos(-1/4)$.
We note that now $B_{\phi}$ denotes a nominal $\pi$ pulse associated with the effective two-photon coupling $\Omega_{\text{eff}}$, the implementation of which requires very large pulse areas, $A_k \gg \pi$ $(k=0,1)$.
%The Raman CP reads
%\be\label{RBB1}
%\text{RBB}1 = \mathbf{B}_{\zeta} \mathbf{B}_{3\zeta} \mathbf{B}_{3\zeta} \mathbf{B}_{\zeta} \mathbf{B}_{0},
%\ee
%with $\mathbf{B}_{\phi}$ denoting the Raman pair $(B^0_{\phi}, B^1_{\phi})$.
The infidelity in this case is plotted in Fig.~\ref{fig:PiPulseVsAreaDetuning} (bottom).
Although the latter sequence contains only 5 pulse pairs, compared to 6 and 10 in the previous ones, it requires much larger total pulse area in order to have effective nominal $\pi$ pulses, and hence this $X$ gate is much slower.

%%%%%%%%%%%%%%%%%%%
\emph{\textbf{Majorana.}}
The $X$ gate can be also produced by using the Majorana decomposition.
As seen from Eq.~\eqref{prop-3s}, if the Cayley-Klein paramaters $a$ and $b$ correspond to complete population transfer in the two-state system ($a=0,|b|=1$), the same will be valid in the Raman system as well.
Hence, we can again use the BB1 composite pulse \eqref{BB1}.
However, each of the $\pi_{\phi_k}$ pulses corresponds to a pair $(\Ra_{\phi_k}, \Rb_{\phi_k}) \equiv \R_{\phi_k}$ in the original basis, where $R=\pi\sqrt{2} (1+\epsilon) = 2Q$.
Explicitly, the Majorana composite Raman X gate reads
\be\label{Majorana-X}
\X_5 = \R_{\zeta} \R_{3\zeta} \R_{3\zeta} \R_{\zeta} \R_{0},
\ee
We note that this composite sequence has the same total pulse area  ($5\times \sqrt{2}\pi$) as the $N=5$ sequence \eqref{Raman-BB3BB3b} used in the MS approach ($2\times 5\times\pi/\sqrt{2}$), and therefore, we can compare the performance of the two methods, see Fig.~\ref{fig:PiPulseVsArea}.
As seen in the figure, by using the CPs, adapted for the Raman qubit, we obtain a robust and high-fidelity $X$ gate in either cases.
We also see that by using the MS approach, we achieve a higher fidelity for the same total pulse area than the Majorana method.

\def\Hadamard{\mathcal{H}}

\begin{figure}[tb]
	\includegraphics[width=7.cm]{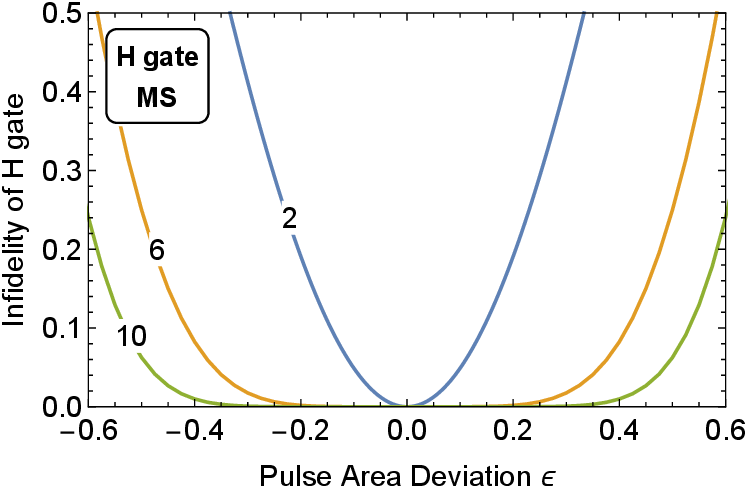}
	\caption{
		Infidelity of the Hadamard gate as a function of the pulse area error for a composite sequence of two, six, and ten pairs of pulses ($N=1,3,5$), see Eq.~\eqref{Raman-BB3BB3-Hadamard}. The curves are the same as those in Fig.~\ref{fig:PiPulseVsArea} (top frame) and are given here only for completeness.
	}
	\label{fig:HadamardVsArea}
\end{figure}

\subsec{Hadamard gate.}
It reads $\frac{1}{\sqrt{2}} (\sigma_x + \sigma_z)$.
%\be\label{HadamardMatrix}
%\Hadamard =
%\frac{1}{\sqrt{2}}\left[ \begin{array}{cc}	1 & 1  \\ 1 & -1 \end{array}\right].
%\ee
%and should not be confused with the Hamiltonian \eqref{Hamiltonian}, also denoted by $\H$.
As seen from Eq.~\eqref{prop-3s}, we cannot generate this gate by using the Majorana decomposition, hence we only use the MS approach.
It follows from Eq.~\eqref{prop-original} that for $\b_0=\sqrt{2+\sqrt{2}}$ and $\b_1=\sqrt{2-\sqrt{2}}$, we have
\be
\U = -\left[ \begin{array}{ccc}
	\frac{1}{\sqrt{2}} &  \frac{1}{\sqrt{2}} & 0 \\
	\frac{1}{\sqrt{2}} & -\frac{1}{\sqrt{2}} & 0 \\
	0 &  0 & 1
\end{array}\right],
\ee
which, up to an irrelevant global sign, is the Hadamard transform of the qubit $\{\ket{0},\ket{1}\}$.
As for the $X$ gate, the MS propagator corresponds to a $2\pi$ pulse, since $\b=2$.
Therefore, we can use the same phases \eqref{phases} to build our composite sequence.
This time, however, $\pi_{\phi_k}$ in the MS basis corresponds to the pulse pair
$[(\xi_0\pi/2)_{\phi_k}, (\xi_1\pi/2)_{\phi_k}] \equiv \S_{\phi_k}$ in the original basis, instead of the $(\Qa_{\phi_k},\Qb_{\phi_k})$ pair used for the $X$ gate. Therefore, we can use the composite sequences for the $X$ gate, and only change the Rabi frequencies.
For example, the $X$ gate CPs \eqref{Raman-BB3BB3} are replaced by
\bse\label{Raman-BB3BB3-Hadamard}
\begin{align}
\mathcal{H}_{6} &= \S_0 \S_{\frac23\pi} \S_0 \S_0 \S_{\frac23\pi} \S_0, \\
\mathcal{H}_{10} &= \S_0 \S_{\frac25\pi} \S_{\frac65\pi} \S_{\frac25\pi} \S_0 \S_0 \S_{\frac25\pi} \S_{\frac65\pi} \S_{\frac25\pi} \S_0,
\end{align}
\ese
which produce composite Raman Hadamard gates.
In Fig.~\ref{fig:HadamardVsArea} we plot the infidelities of the composite Hadamard gates, produced by sequences, consisting of 2, 6, and 10 pairs of pulses.

In the presence of detuning, one can proceed in the same way as for the $X$ gate.
The $2\pi$ pulse, which produces a propagator with $a=-1$, is replaced by a phase gate, which produces $a=-\e^{\i\delta}$, or alternatively, a universal CP can be used, in the case of moderate detuning.
For large detuning, again adiabatic elimination and the half-$\pi$ BB1 pulse \cite{Wimperis1994} can be used.

%=================================================================
%=================================================================

\subsec{Rotation gate.}
In order to produce composite rotation gates, we proceed in a way similar to the $X$ and Hadamard gates.
Let us set $\b_{0}=2\sin(\theta/2)$ and $\b_{1}=-2\cos(\theta/2)$.
This choice produces a $2\pi$ pulse in the MS basis, just as before.
Then the propagator reads
\be
\U= \left[ \begin{array}{ccc}
	\cos\theta & \sin\theta & 0 \\
	\sin\theta &  -\cos\theta & 0 \\
	0 &  0 & -1
\end{array}\right].
\ee
This propagator describes a qubit rotation, although not in the usual form $e^{i \theta \sigma_y}$.
As an example, for a robust $\pi/3$ rotation, we use a composite sequence with the same phases as in the previous two subsections, but each MS $\pi_{\phi_k}$ pulse is produced by the pair $[(\pi/2)^0_{\phi_k}, (-\pi\sqrt{3}/2)^1_{\phi_k}]$ in the original basis.
It can be shown that the infidelity of the rotation gate does not depend on the angle $\theta$ and is therefore the same as the infidelity of the $X$ and Hadamard gates.
This is because these can be considered as rotations with $\theta=\pi/2$ and $\pi/4$, respectively. Even more, an analytic formula for the infidelity of the $X$, Hadamard, and rotation gates can be derived,
\be
D = 2\sin^{2N}\left(\frac{\pi\epsilon}{2}\right),
\ee
which demonstrates that the robustness of each of the composite gates, produced by the MS approach, is of the order $O(\epsilon^{2N})$.

%=================================================================
\begin{figure}[tb]
	\includegraphics[width=7.cm]{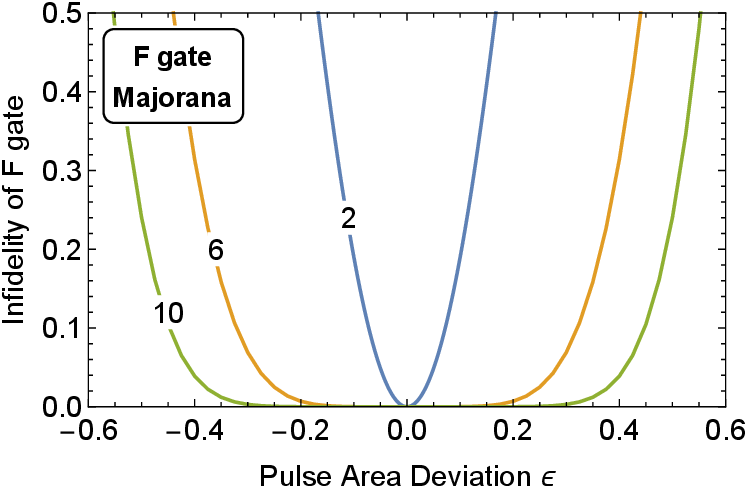}
	\caption{
		Infidelity of the phase shift gate as a function of the pulse area error for a composite sequence of two, six, and ten pairs of pulses ($N=1,3,5$) and $\eta=\pi/4$. The explicit pulse sequence for $N=3$ is given by Eq.~\eqref{Raman-phase gate}.
	}
	\label{fig:PhaseGateVsArea}
\end{figure}
%=================================================================

\subsec{Phase gate.}
It reads $ \hat{F} = \exp[i \eta \hat{\sigma}_z/2] $.
The composite version of this gate cannot be produced by the MS approach but only by the Majorana decomposition.
We notice that if the propagator \eqref{prop-2s} is a phase gate with a phase $\eta/2$, then the propagator \eqref{prop-3s} is a phase gate with a phase $\eta$.
Therefore, in order to produce a composite Raman phase gate we can use the available two-state CPs.
A number of composite phase gates have been presented in Ref.~\cite{PhaseGate} and we can directly implement them for the Raman qubit following the above argument.
As a specific example, the sequence of two three-pulse $\pi$ CPs $\B_3(0)\B_3(\eta/2)$, i.e.,
\be\label{Raman-phase gate}
\F_{6} = \R_0 \R_{\frac23\pi} \R_0 \R_{\frac12\eta} \R_{\frac23\pi+\frac12\eta} \R_{\frac12\eta},
\ee
produces a composite Raman phase gate of phase $\eta$.
This approach can be applied for arbitrarily long sequences by using longer BB CPs of Eq.~\eqref{BBCP} with the phases \eqref{phases}.
In Fig.~\ref{fig:PhaseGateVsArea} the infidelity of these sequences up to $N=5$ is plotted for $\eta=\pi/4$.
This value of $\eta$ corresponds to the $T$ gate, which is widely used in quantum computing \cite{QIP}.
As seen from the figure, composite pulses implement robust and high-fidelity phase gates.

\subsec{Discussion and Conclusions. }%\label{Sec-Conclusions}
In this work, we developed a systematic framework for creating robust and high-fidelity quantum gates in Raman qubits.
Our approach uses composite sequences of pulse pairs and is based on two transformations, the Morris-Shore transformation and the Majorana decomposition.
These allow the three-state Raman problem to be treated as a two-state system and hence to benefit from the vast amount of broadband composite pulses developed for simple two-state systems.
We have constructed and numerically demonstrated the $X$, Hadamard, rotation, and phase (including the $S$, $T$, and $Z$ gates) gates.
The $X$, Hadamard, and rotation gates, in particular, are constructed in the same manner, using composite sequences with the same phases and the same RMS pulse area of $2\pi$ of each pulse pair, but different ratios of the Raman couplings.
Implementations for both on-resonance and arbitrarily detuned (small, medium, and large) ancilla middle state are presented.
One could easily apply this general approach to other commonly used gates like the $Y$ gate.
The proposed composite Raman gates can allow one to implement single-qubit operations at ultrahigh fidelity and resilient to experimental errors, as required for efficient quantum computation.

\acknowledgments
This work is supported by the European Commission's Horizon-2020 Flagship on Quantum Technologies project 820314 (MicroQC).

%%%%%%%%%%%%%%%%%%%%%%%%%%%%%%%%%%%%%%%%%%%%%%%%%%%%%%%%%%%%%%%%%%%%%%%%%%%%%%%%%%%%%%%%%%%%%%%%%
%%%%%%%%%%%%%%%%%%%%%%%%%%%%%%%%%%%%%%%%%%%%%%%%%%%%%%%%%%%%%%%%%%%%%%%%%%%%%%%%%%%%%%%%%%%%%%%%%
%%%%%%%%%%%%%%%%%%%%%%%%%%%%%%%%%%%%%%%%%%%%%%%%%%%%%%%%%%%%%%%%%%%%%%%%%%%%%%%%%%%%%%%%%%%%%%%%%
%%%%%%%%%%%%%%%%%%%%%%%%%%%%%%%%%%%%%%%%%%%%%%%%%%%%%%%%%%%%%%%%%%%%%%%%%%%%%%%%%%%%%%%%%%%%%%%%%
%%%%%%%%%%%%%%%%%%%%%%%%%%%%%%%%%%%%%%%%%%%%%%%%%%%%%%%%%%%%%%%%%%%%%%%%%%%%%%%%%%%%%%%%%%%%%%%%%

\end{document}